\documentclass[final,5p,times,twocolumn]{elsarticle} 
\biboptions{sort&compress}

\usepackage{amssymb}
\usepackage{amsthm}
\usepackage{graphicx} 
\usepackage{caption}
\usepackage[colorlinks,linkcolor=red,anchorcolor=blue,citecolor=green]{hyperref}
\usepackage{courier}
\usepackage[switch]{lineno} 

\usepackage{listings}
\usepackage{framed} 
\usepackage{listings}
\usepackage{xcolor}
\definecolor{codegreen}{rgb}{0,0.6,0}
\definecolor{codegray}{rgb}{0.5,0.5,0.5}
\definecolor{codepurple}{rgb}{0.58,0,0.82}
\definecolor{backcolour}{rgb}{0.95,0.95,0.95}
\definecolor{black}{rgb}{0.0,0.0,0.0}

\lstdefinestyle{mystyle}{
    backgroundcolor=\color{backcolour},   
    commentstyle=\color{codegreen},
    keywordstyle=\color{magenta},
    numberstyle=\tiny\color{codegray},
    stringstyle=\color{codepurple},
    basicstyle=\ttfamily\footnotesize,
    breakatwhitespace=false,         
    breaklines=true,                 
    captionpos=b,                    
    keepspaces=true,                 
    numbers=left,                    
    numbersep=5pt,                  
    showspaces=false,                
    showstringspaces=false,
    showtabs=false,                  
    tabsize=2,
    frame=shadowbox,
    rulecolor=\color{black},
}
\newcounter{bla}

\journal{Computer Physics Communications}

\begin{document}
\captionsetup[figure]{labelfont={bf},labelformat={default},labelsep=period,name={Fig.}}
\begin{frontmatter}



\title{FF7: A Code Package for High-throughput Calculations and Constructing Materials Database}


\author[a]{Tiancheng Ma}
\author[a]{Zihan Zhang}
\author[a]{Shuting Wu}
\author[a]{Defang Duan\corref{author}}
\author[b,a]{Tian Cui}

\cortext[author] {Corresponding author.\\\textit{E-mail address:} duandf@jlu.edu.cn}
\address[a]{Key Laboratory of Material Simulation Methods $\&$ Software of Ministry of Education and State Key Laboratory of Superhard Materials, College of Physics, Jilin University, Changchun 130012, China}
\address[b]{Institute of High Pressure Physics, School of Physical Science and Technology, Ningbo University, Ningbo 315211, China}

\begin{abstract}
Decades accumulation of theory simulations lead to boom in material database, which combined with machine learning methods has been a valuable driver for the data-intensive material discovery, i.e., the fourth research paradigm. 
However, construction of segmented databases and data reuse in generic databases with uniform parameters still lack easy-to-use code tools. 
We herein develop a code package named FF7 (Fast Funnel with 7 modules) to provide command-line based interactive interface for performing customized high-throughput calculations and building your own handy databases. 
Data correlation studies and material property prediction can progress by built-in installation-free artificial neural network module and various post processing functions are also supported by auxiliary module. 
This paper shows the usage of FF7 code package and demonstrates its usefulness by example of database driven thermodynamic stability high-throughput calculation and machine learning model for predicting the superconducting critical temperature of clathrate hydrides. 
\end{abstract}

\begin{keyword}
high-throughput calculation; material database; mechine learning.

\end{keyword}

\end{frontmatter}


\newpage
\noindent
{\bf PROGRAM SUMMARY}

\begin{small}
\noindent
{\em Program Title:} FF7 \\
{\em CPC Library link to program files:} (to be added by Technical Editor) \\
{\em Code Ocean capsule:} (to be added by Technical Editor)\\
{\em Licensing provisions:} MIT  \\
{\em Programming language:} Python                  \\
{\em Nature of problem:} Since the data-intensive material discovery under the fourth paradigm progresses by the boom of database and machine learning, 
a handy code tool for performing flexible high-throughput theory simulations, 
building your own database for specific research interest and constructing artificial neural network model for material properties prediction is highly desired.
Comprehensive post-processing and graphic drawing functions are also required. \\
{\em Solution method:} The first principal density function theory (DFT) simulations are performed by the VASP and Quantum Espresso code package. For flexibility of high-throughput calculations and database construction,
the calculation tasks are abstracted into a “calculation card”. It contains the DFT software name, the computational parameters, the variant (or file) to be written to the database, etc., all of which can be customized by users.
The functions of FF7 code package are realized via the Linux command line for ease of use.\\
{\em Additional comments including restrictions and unusual features:} This program works on the Linux operating system with VASP and Quantum Espresso code packages installed.\\
   \\


\end{small}

\section{Introduction}
\label{}
Data-intensive scientific discovery was first proposed by Jim Gray in 2007 as the fourth research 
paradigm, following the traditional empirical trial-and-error method, theoretical modelling approaches and 
software simulation\cite{fourthpar}. Its research idea is analogous to that in the discovery of the laws of planetary motion by 
Tycho Brahe and his assistant Johannes Kepler where the creation of theories is driven by the mining and 
analysis of captured and carefully archived massive astronomical observation data. It emphasizes that big data 
and statistical learning methods, or machine learning methods, are the primarily basis of the fourth research 
paradigm\cite{prl_bigdata}. Thanks to the development of material simulation algorithms and first-principles DFT calculation 
software (e.g. VASP\cite{VASP}, CASTEP\cite{CASTEP} and Quantum Espresso(QE)\cite{QE}), vast volumes of raw materials science data have been accumulated by the high-performance
computing resources on a 24/7 basis and has brought us to the stage of transformation of materials 
research methods. At the same time, the boom in machine learning models, like CGCNN\cite{CGCNN} and ALIGNN\cite{ALIGNN} which targets on crystal represention, has also become another push for 
us to move forward. The stage is set for the fourth paradigm to be applied in material design field. In 2011, the 
Materials Genome Initiative (MGI) was proposed and take the lead in bringing related research to the 
fore\cite{MGI}. It shifts our emphasis to targeted materials discovery via high-throughput identification of the key factors 
(i.e., “genes”) and via showing how these factors can be quantitatively integrated by statistical learning 
methods into design rules (i.e., “gene sequencing”) governing targeted materials functionality. The MGI 
 generally involves three basic data activities: capture, curation, and analysis, which correspond to three 
infrastructures of high-throughput DFT calculation tools, databases, and machine learning models, respectively. 
Relying on the materials analysis python library pymatgen\cite{pymatgen_htp}, the largest inorganic crystal materials database, 
Materials Project, containing hundred thousand items of the structural (e.g. lattice parameters, space group) 
and material properties information (e.g. electronic band structure, phonon dispersion curves and elastic 
tensors), was created and became one of the key infrastructures in the materials discovery field\cite{MP}. Based on the 
rise of various databases such as Materials Project\cite{MP}, ICSD\cite{ICSD}, OQMD\cite{OQMD}, etc. and the rapid development of machine 
learning arithmetic represented by the introduction of Graph Convolutional Networks, a series of machine learning 
models based on crystal geometric features or simple descriptors for predicting material properties (e.g. 
formation enthalpy\cite{CGCNN,ALIGNN}, bandgap\cite{CGCNN,JAMIP}, hardness\cite{hardness} and superconducting transition temperatures ($T_c$)\cite{PRM_mlTc,PRB2023_mlTc,Hutcheon,Shipley}) have been developed, 
which have significantly reduced computational costs and facilitated the subsequent discovery of 
materials with target properties. 

In addition to utilizing large scale general-purpose databases, the development of dedicated databases by 
individuals to accelerate discovery of materials with specific properties is also a way forward. Although the 
data scale may not be as large as these general databases, it can also play an important role in supporting the 
data-intensive discovery of targeted materials by its localization and accuracy. Successful 
examples in bandgap prediction\cite{CGCNN,JAMIP}, high entropy alloy design\cite{sa_luo} and $T_c$ prediction of 
superconducting materials validate the feasibility of this idea. The 
need for fine-grained domain databases, spawned by the MGI and unmet by general-purpose databases, 
creates an urgent requirement for user-friendly, full featured code tools that can assist in building one’s own 
database. Furthermore, reuse of data in general-purpose databases is a major problem, especially for formation 
enthalpy convex hull calculations where uniform parameters are highly demanded, and this requires code tools 
for high-throughput calculations and databases construction. Finally, the stock of data dispersed among small 
laboratories or individuals cannot be ignored. Based on the MGI’s principle of data inclusiveness, code 
tools are needed for interface-unified database construction, allowing anyone to contribute to the data accumulation.

We have many excellent tools to assist DFT high-throughput computation, such as VASPKIT\cite{VASPKIT}, qvasp\cite{QVASP}, JAMIP\cite{JAMIP}, VASPMATE\cite{VASPMATE}, etc., 
but there is still a lack of a code tool that can be tightly connected to high-throughput computation, 
create databases senselessly, and provide powerful database access and management functions.
We herein report a code package named FF7 (Fast Funnels with 7 modules) that fully meet the 
requirements of constructing private database through high-throughput (HTP) calculations and support the 
workflow of data-intensive materials discovery. The mainstream density function theory (DFT) calculation 
software VASP and QE are both supported, and notably latter’s high-throughput computational tools are 
developed for the first time. Several built-in HTP calculation functions could help to build the basic database 
including stoichiometry, structure, energy, etc. and the code architecture design strategy of “calculation card” 
allows the highly customized HTP calculations by users and constructing databases for niche areas. 
Furthermore, full featured post-processing tools and artificial neural network module for materials properties 
analysis and predictions are also integrated in it. All functions of the FF7 code package are available through 
the Linux command-line based user-friendly command with a uniform command style for two DFT software. 
The code architecture (section 2), detailed usages for each module (section 3) and example of aiding data 
driven materials discovery (section 4) are discussed in the following sections. \\

\begin{figure}[htbp]
  \centerline{\includegraphics[width=1\linewidth]{"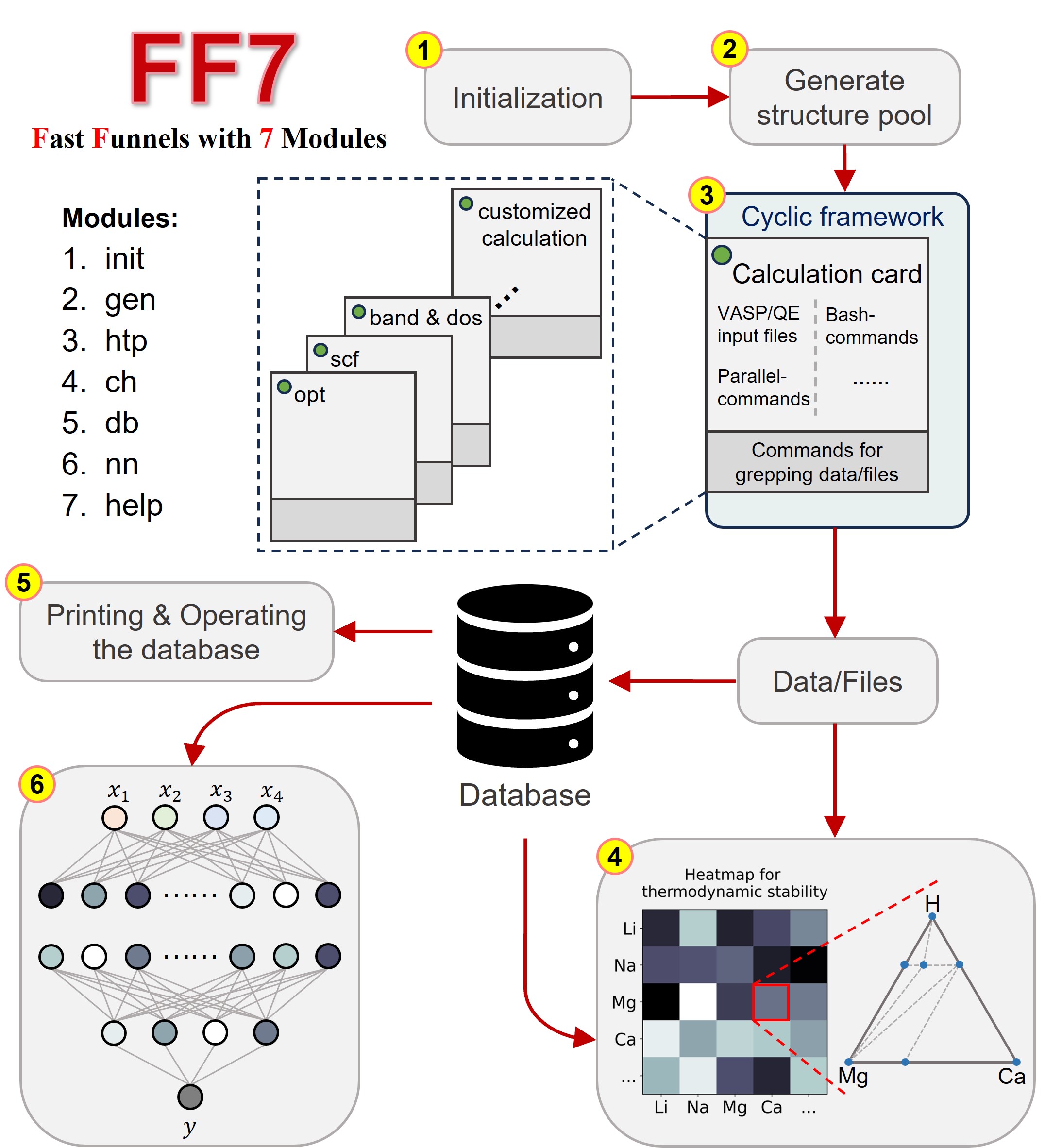"}}
  \caption{Architecture of FF7 code package.}
  \label{arch}
\end{figure}


\section{Architecture and design strategy}
\label{}
The FF7 code package consists of seven main modules, which are streamed together in a workflow that fits the fourth research paradigm, as shown in Fig. \ref{arch}.
Firstly, the ``init'' module initialize the FF7 code package by loading basic parameters such as the DFT software path, database path, pseudopotentials path, etc.
Then, a structure pool, consisting of a series of crystal files in POSCAR format with ``.vasp'' suffix, is provided by user or generated by the ``gen'' module based on different strategies. 
The ``htp'' module can help to perform HTP calculations traversing it, in which the DFT calculations is supported by the mainstream software of VASP and QE. 
In ``htp'' module, the high-throughput calculation workflow is abstracted into a ``calculation card'' in a cyclic framework, as shown in the third step in Fig. \ref{arch}.
The ``calculation card'' controls the DFT calculations and the rules for creating database, e.g. which variable or file will be stored in the database.
Apart from a wealth of built-in calculation cards in FF7, users can customize ``calculation cards'' to perform task-specific high-throughput calculations and create databases that match their own research interests, which greatly ensures the flexibility of FF7 code package.
The post-processing functions are also equipped for most of the built-in HTP calculations by ``post'' module, including extracting and summarizing data from output files, graphic drawing, and as the highlight, the database driven high-throughput calculations for thermodynamic stability (see the fourth step in Fig. \ref{arch}).
The ``db'' module allows users to print the constructed database on the screen and supplied command-line based functions of adding, deleting and extracting data of it.
Finally, the ``nn'' module and self-built databases are utilized for the construction and training of artificial neural networks to have the role of predicting material properties, which in turn facilitates material discovery, realizing the research life cycle in the fourth research paradigm. 
Overall, the design strategy of FF7 code packages targeted on the ease to use and flexibility. 
The database occupies the most important place, and all the modules are highly interconnected with it, which is one of the superiorities of the FF7 code package. \\

\section{Functions and usage}
\label{FaU}
We show the functions and usages of FF7 code packages in this section.
In generally, the functions in FF7 are implemented by command line on the Linux system with a uniform style of \\
\texttt{\$ ff7 module func -para1 xx -para2 xx …} ,\\
where the ``\texttt{module}'', ``\texttt{func}'' and ``\texttt{para}'' denote the module name, function of the module and parameters for the function, respectively.
Detailed discussions for each module are listed below.

\subsection{\textbf{init}}
\label{init}
The ``init'' module is executed automatically at the beginning of the FF7 program life circle
and users have to complete the initialization file before running FF7 code package.
It loads the default variables  including the path of the DFT calculation software VASP and QE, the pseudopotentials path and server configurations 
from the initialization file “/.../installation/init/BASE.ini” (the “installation” denotes the installation path of the FF7 code package).
As the VASP provides a complete pseudopotentials package with different pseudopotential versions for each element, users can specify the pseudopotentials name by editing the file “/.../installation/init/POTCAR.ini”.
For the QE, the pseudopotentials directory needs to be created by users through collecting pseudopotential files for each element with name of “element.UPF”.
Users can specify three pseudopotential paths for QE according to the pseudopotential type of US, PAW and NC respectively. 

\subsection{\textbf{gen}}
\label{gen}
The structure pool is a directory containing structure files with suffix “.vasp” in the form of POSCAR, which is the main work path of FF7 code package.
Users can easily build their structure pools by copying structures of interest to them for high-throughput calculations.
Also, the “gen” module can help to construct a structure pool through element substitution based on the POSCAR-formatted structure file “seed”.
For binary compounds, the replacement of the first element in the “seed” file can be done by with the command\\
\texttt{ff7 gen -1 [Li,Na,K,Rb,Cs].}\\
The “gen” module also supports the structure generation of ternary compounds from “seed” file. Commands for replacing two spatial unequal and equivalent elements are shown line 1 and line 11 in Fig. \ref{scr}, respectively, 
and the FF7 will print a brief for the generated compounds on the screen for checking (see line 2-9 and 12-27 in Fig. \ref{scr}).

\begin{figure}[htbp]
\begin{lstlisting}[language=bash]
$ ff7 gen -1 [Li,Na,K,Rb] -2 [Be,Si,P]
   M_1 X_1 H_8
    | Be Si P
  -----------
  Li| @  @  @
X Na| @  @  @ 
  K | @  @  @
  Rb| @  @  @ 
        M

$ ff7 gen -1 -2 [Mg,Ca,Sc,Ti,Sr,Y,Zr,Ba,La,Ce,Hf,Th]
                M_1 X_1 H_12
    | Mg Ca Sc Ti Sr Y  Zr Ba La Ce Hf Th
  --+------------------------------------
  Mg|
  Ca| @ 
  Sc| @  @ 
  Ti| @  @  @ 
  Sr| @  @  @  @ 
X Y | @  @  @  @  @ 
  Zr| @  @  @  @  @  @ 
  Ba| @  @  @  @  @  @  @ 
  La| @  @  @  @  @  @  @  @ 
  Ce| @  @  @  @  @  @  @  @  @ 
  Hf| @  @  @  @  @  @  @  @  @  @ 
  Th| @  @  @  @  @  @  @  @  @  @  @ 
                     M
\end{lstlisting}
  \caption{Commands and output of the ``gen'' module. }
  \label{scr}
\end{figure}

\subsection{\textbf{htp}}
\label{htp}
\begin{figure}[htbp]
  \centerline{\includegraphics[width=0.65\linewidth]{"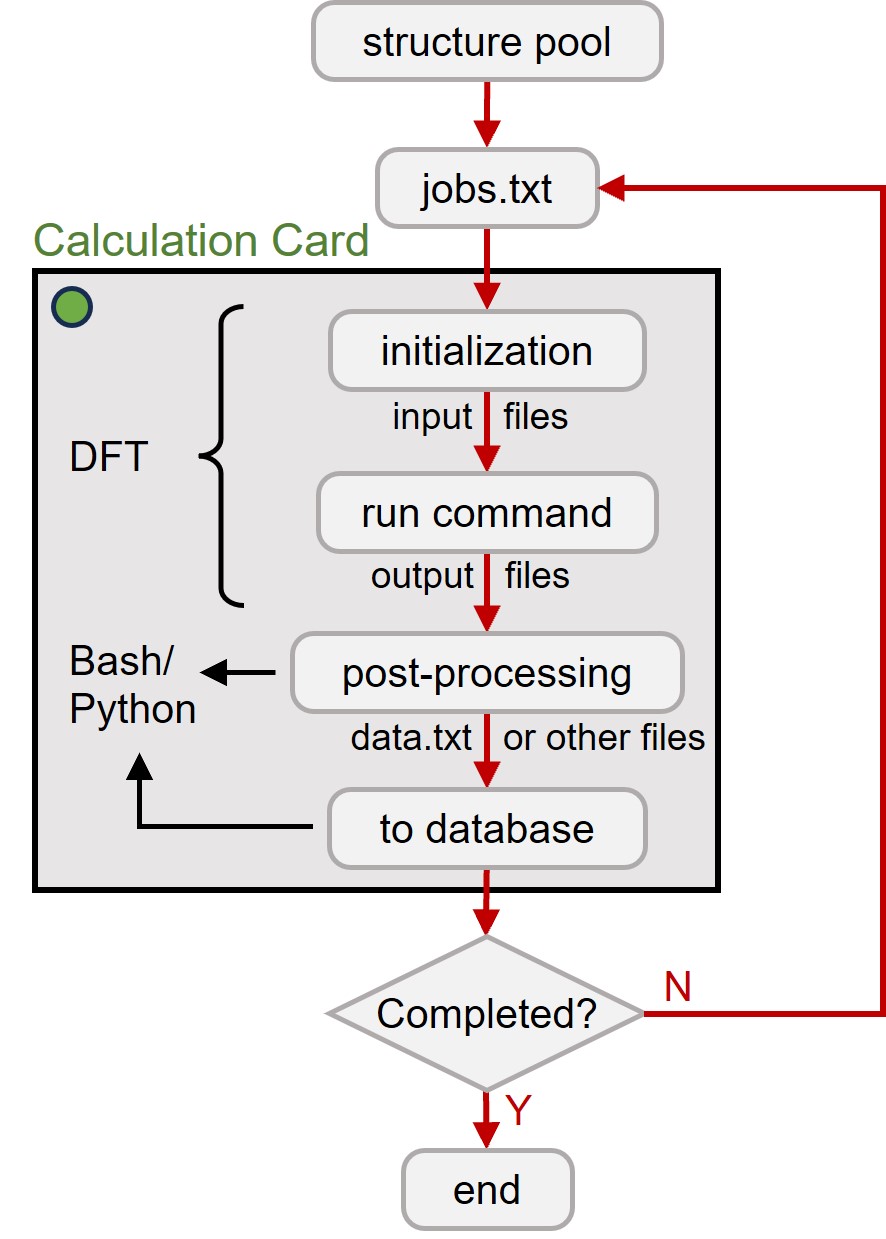"}}
  \caption{The workflow of ``htp'' module and the calculation card.}
  \label{fhtp}
\end{figure}

The “htp” module facilitates the high-throughput DFT calculations for the compounds in the structure 
pool and transfers data or file results, which vary depending on the calculation task, to the database. The 
workflow of the “htp” module is shown in Fig. \ref{fhtp}. First, it renames the compounds in structure pool in the form 
of “ele1$\_$x1$\_$ele2$\_$x2$\_$sg.vasp”, where “ele”, “x” and “sg” denote the element, stoichiometric number and space 
group number respectively, and lists them in the file “jobs.txt” as a queue for high-throughput calculations. 
The calculation task for each compound can be abstracted as a calculation card (see Fig. \ref{fhtp}) that records how 
the input files are generated, the commands to perform DFT calculation, the post-processing codes for output 
files and how the calculated results are transferred to the database. 
The FF7 code package provides several 
calculation cards for various DFT calculations by VASP and QE. For example, users can perform structural 
optimizations for all compounds with VASP by command \\
\texttt{\$ ff7 htp opt -p 200 -e 600 -k 0.03} ,\\
where the “-p”, “-e” and “-k” denote the parameters of pressure (GPa, default as 0.001), energy cutoff (eV) 
and k points mesh spacing (2$\pi/\AA$, default as 0.03), respectively. The command only needs to be changed slightly as\\
\texttt{\$ ff7 htp opt\_qe -p 200 -e 80 -pps us}\\
for the QE supported DFT calculations. It should be noted that the unit of energy cutoff here is R.y and users 
need to specify the pseudopotentials type by parameter “-pps” (default as “us”). After the structural optimization, 
the entries containing stoichiometry, crystal structure, space group symbol and energy will be automatically 
transferred to the database forming the basic framework of the database. We therefore recommend running 
“opt” function first to build the initial database. The FF7 code package also provides built-in calculation cards 
of “scf” for static self-consistent calculations, “bandos” for electronic band structures and density of states and 
“elf” for electron localization functions. The DFT calculations for each compound are performed under the 
path of “/.../structure$\_$pool/compound$\_$name/function” and the calculated values or files are automatically 
stored in the database. For numeric results, they are stored directly in the database, while for file results, the 
files are copied to a subfolder and the path will be recorded by database. For every function, the “-db” 
parameter (default as “/.../installation/db”) can specify the database path. Users can also tell the FF7 to only 
perform high-throughput calculations without any connection to the database with the flag “-nodb”. All of 
these functions support DFT calculations by VASP and QE, with the difference being the suffix “$\_$qe” to the 
function name. The commands, for the function “bandos” as an example, are\\
\texttt{\$ ff7 htp bandos -e 600 -k 0.03}\\
and \\
\texttt{\$ ff7 htp bandos\_qe -e 80 -k 0.03}\\
for VASP and QE, respectively. Calculations combined with structural optimization and electronic band 
structure are supported by adding the flag “-dopt”:\\
\texttt{\$ ff7 htp bandos -e 600 -dopt -p 200.}\\
The FF7 code package supports the phonon calculation by density functional perturbation theory and finite 
displacement method using VASP+phonopy and QE, respectively, and the example commands for phonon 
calculation combined with structural optimization are\\
\texttt{\$ ff7 htp phonon -e 600 -k 0.02 -dim 2 2 2 -dopt -p 200}\\
and \\
\texttt{\$ ff7 htp phonon\_qe -e 80 -k 3 3 3 -q 12 12 12 -pps us -dopt -p 200}.\\
The “dim” parameter is the rules for creating supercells from unit cell that corresponds to the “-dim”
tags in phonopy. When calculating the phonon spectra with QE, the electron-phonon coupling is 
calculated at the same time, which does not take much extra time. 

\begin{figure}[htbp]
\begin{lstlisting}[language=bash]
Software        : vasp
Dirname         : fermi
InFile          : INCAR.self
RunCommand      : mpirun -np 8 /../vasp/bin/vasp_std
KPOINTS         : 0.03
# KPOINTS       : HIGH_SYMMETRY_PATH
DataType        : file
DataLabel       : Efermi_file
GrepDataCommand : grep 'E-fermi' OUTCAR | awk '{print $3}' > Efermi_file.txt
# GrepDatafile  : grep.py

\end{lstlisting}
  \caption{Input files templet for customizing calculation cards.}
  \label{cc}
\end{figure}

In fact, the built-in functionality is far from satisfying all users’ requirements. 
As a solution, users can perform high-throughput calculation by customize the calculation cards for their various research interest by the command\\
\texttt{\$ ff7 htp self -file self.in.}\\
The templet of file “self.ini” that governs the high-throughput DFT calculation and constructing database is shown in Fig. \ref{cc} and 
each component of this file are described as follows:\\
\begin{itemize}
  \item[]
(1) Software name, ``vasp'' and ``qe'' are optional;\\
(2) Directory name where calculations are performed; \\
(3) Input file name; \\
(4) Command to run DFT calculations; \\
(5) K point mesh spacing valure; \\
(6) Generating k-points along high symmetry path; \\
(7) Type of the results, “file” and “value” are optional; \\
(8) The label of the result to be stored in the database; \\
(9) The command to extract the result data which must be stored in a file named “DataLabel.txt”; \\
(10) The python code is also support to grape the result data which must be stored in a file named “DataLabel.txt”.\\
\end{itemize}
Overall, the design strategy of highly customizable calculation cards gives users more flexibility for high-throughput calculations and constructing their own database.

\renewcommand{\dblfloatpagefraction}{.9}
\begin{figure*}[htbp]
  \centerline{\includegraphics[width=1\linewidth]{"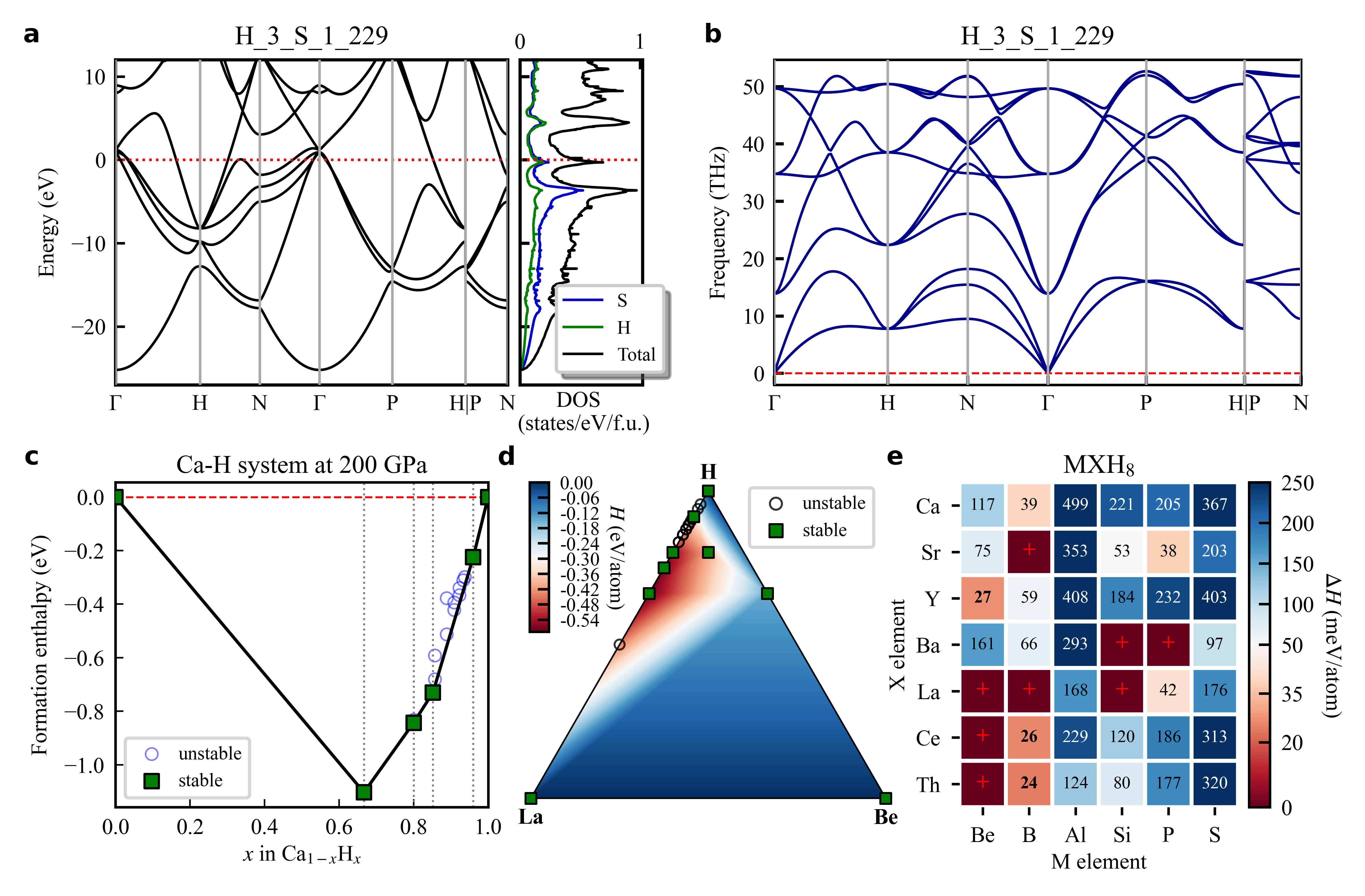"}}
  \caption{(a) Electronic band structures and density of states and (b) phonon spectrum of H3S drawn by “post” module. 
  (c) Formation enthalpy convex hull of Ca-H system and (d) La-Be-H system at 200 GPa generated by “post” module. 
  (e) Thermodynamic stability heatmap diagram of hydrides MXH$_8$ (M=Be, B, Al, Si, P, S, X=Ca, Sr, Y, Ba, La, Ce, Th) calculated and drawn by ff7 code package combined with self-build high pressure database at 200 GPa. 
  Colors in (d) and (e) denote formation enthalpy and enthalpy above the convex hull, respectively. 
  Red crosses in (e) represent compounds with thermodynamic stability. }
  \label{post_cv}
\end{figure*}

\subsection{\textbf{post}}
\label{post}

The “post” module mainly implements post-processing functions such as drawing diagrams and 
generating formation enthalpy convex hulls. After high-throughput calculations for the electronic band 
structures and density of states by the “htp” module, users can run command\\
\texttt{\$ ff7 post bandos -lim -27 8}\\
in the path “/.../structure$\_$pool/compound$\_$name” to draw an electronic band structure diagram for a certain 
compound, where the parameter “-lim” controls the energy limit. In addition, the command\\
\texttt{\$ ff7 post bandos -byjobs}\\
ran in “/structure$\_$pool” directory allows batch drawing for compounds listed in the “jobs.txt” file. In keeping 
with the consistent unity of command style, the command with addition of “$\_$qe” to function can processes the 
results calculated by QE. Similar drawing commands also support the visualization of phonon spectra, except 
that the function “bandos” is replaced by “phonon”. Taking H$_3$S\cite{NP_H3S, N_H3S, ddfH3S} as an example, the electronic band 
structures and phonon spectra drawn by FF7 are shown in Fig. \ref{post_cv}a-b. The “post” module provides full functions 
for generating formation enthalpy convex hull based on the self-built database. Users can use command\\
\texttt{\$ ff7 post ch -path /your/path -nodb}\\
to generate the convex hull for the compounds in the path declared by “-path” parameter (default as “./”), 
where the format of compounds file should follow that in “opt” calculations. To generate an enthalpy convex 
based on a self-constructed database, users need to specify the path of the database:\\
\texttt{\$ ff7 post ch -dbpath /path/to/self.db} .\\
In addition, users can create formation enthalpy convex hull diagrams based on the elements directly from a 
self-built database: \\
\texttt{\$ ff7 post ch -dbpath self.db -sys H S} .\\
The convex hull diagram for Ca-H\cite{pnas_CaH6, an_Ca4H23,ma_CaH6} binary and La-Be-H\cite{exp_LaBeH8, pre_LaBeH8} ternary system drawn by FF7 is shown in Fig. \ref{post_cv}c-d. Although 
FF7 does not support the drawing of higher dimensional convex hulls, it supports the creation of them and 
could print them on the screen. For the high-throughput calculations for ternary compounds based on the 
elemental substitution with “opt” function, the “post” module can create convex hulls for the compounds in 
file “jobs.txt” and summarizes the thermodynamic stability information into a heatmap diagram, as shown in 
Fig. \ref{post_cv}e, and the command is \\
\texttt{\$ ff7 post heatmap -M 1 -X 2},\\
where the “-M” and “-X” denote the index of substituted element in the file “seed”. Additionally, the FF7 code 
package provide the interface with the database of material project (MP). Users need to construct a database for 
single substances in MP and run the command\\
\texttt{\$ ff7 post ch -path ./ -MPsingleDb /path/to/MPsingle.db -MPID xxx}, \\
where the “MPID” is the ID of the Materials Project account for visiting the online database API. Before that, 
you need to build a MP single substance database and declare it with the “-MPsingleDb” parameter. The single 
substance database is used to calculate the formation enthalpy and that for binary and ternary compounds are 
directly caught through the API of the material projects. This strategy for creating convex hulls save a lot of 
computational effort while guaranteeing accuracy.

\subsection{\textbf{db}}
\label{db}

The “db” module is a main module in FF7 code package with database engine being the self-contained 
and highly reliable sqlite3 Python package. It provides a full-featured command line interface for browsing, 
manipulating and outputting self-built databases, which correspond to the “show”, “add” (or “delete”) and
“save” (or “catch”) function, respectively. The database accepts numeric type and file type data. Numeric data 
are directly stored in the database while the file type data are copied to the subfolder in the database path and 
the database record the storage path. User can use the command \\
\texttt{\$ ff7 db show -summary}\\
to preview the important information of database including the database path, the columns names and the 
number of database entries. Also, command\\
\texttt{\$ ff7 db show -summary -dbpath /your/db/self.db}\\
allows users to print all the data of a specific database that declared by the “-dbpath” parameter. The “-
compound” and “-cols” parameters can completely locate a data or a file in the database, and users can take a 
full control of the database via the “add” (or “update”) and “delete” functions. The example commands for  
deleting and adding data are  \\
\texttt{\$ ff7 db add -compound H3S -cols bands } \\
and \\
\texttt{\$ ff7 db add -compound H3S -cols energy -data 10.68}, \\respectively. 
The “db” module also supports data retrieval by element system and column name, which are accepted by the 
“-system” and “-cols” parameters respectively. The two parameters can be used together to specify the data to 
be manipulated. For example, to retrieve the energy of the compounds in the S-H system in the database, users 
can print the retrieved information on the screen with the command \\
\texttt{\$ ff7 db show -system S H -cols energy.}\\
Anything printed on the screen can be stored in a file by replacing the “show” function with the “save” function. 
Furthermore, to obtain the file type data, users need to use the “catch” function with the command\\
\texttt{\$ ff7 db catch -system S H -cols bands} \\
to copy the files to the current folder. The “db” module for the creation and modification of the database 
provides a more flexible way to greatly reduce the threshold for constructing database and provide a rich of 
interface to the database making it accessible to users and allow the interactions with other module in FF7 code 
package and can be merged in users’ familiar workflow easier. \\

\renewcommand{\dblfloatpagefraction}{.9}
\begin{figure*}[htbp]
  \includegraphics[width=1\linewidth]{"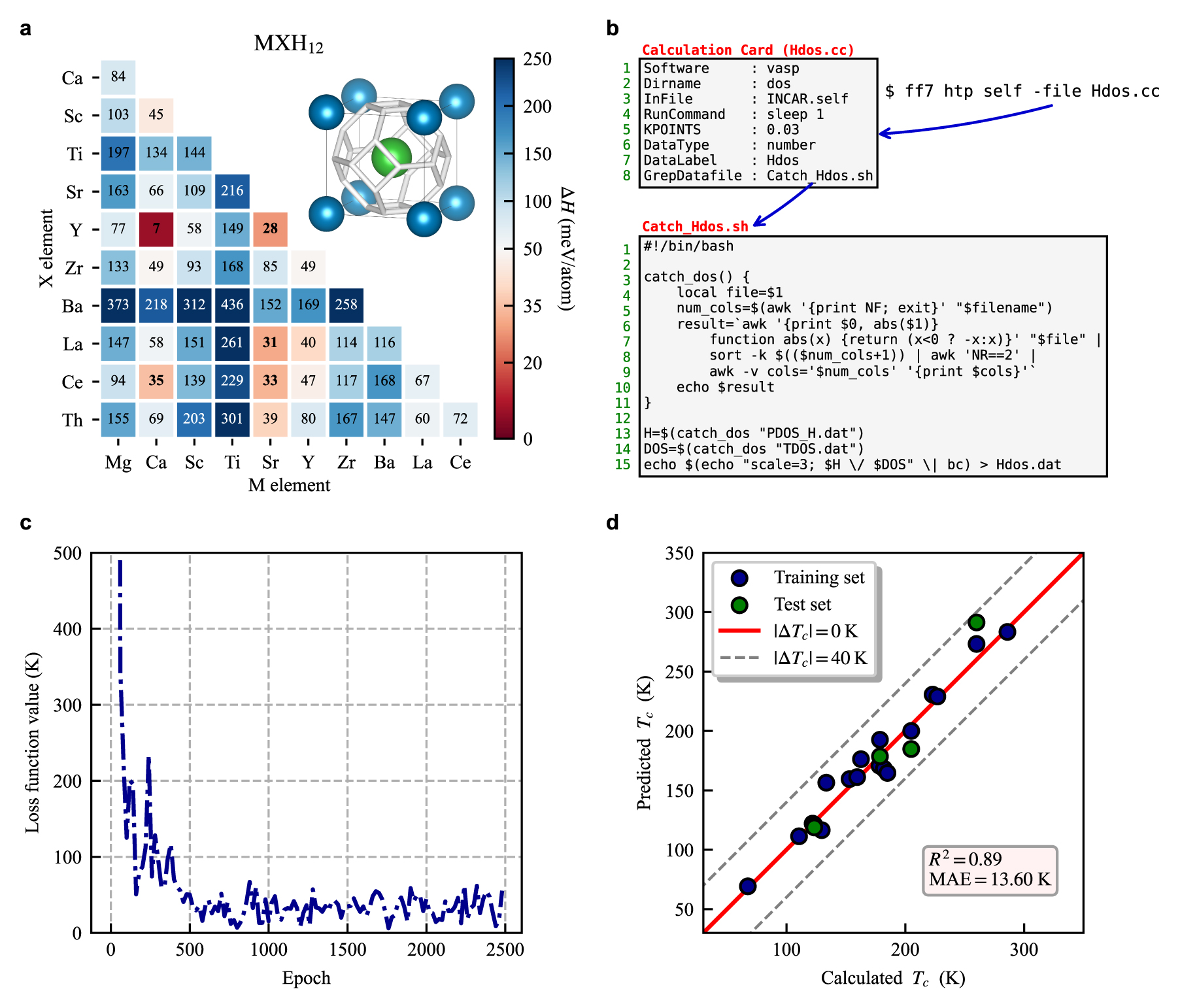"}
  \caption{(a) The formation enthalpy above the convex hull of compound MXH$_{12}$ at 200 GPa. 
           (b) The "Hdos.cc" file and "Catch\_Hdos.sh" file. 
           (c) The loss function value (MAE) for predicting Tc of clathrate hydrides as the function of train epoch. 
           (d) The machine learning model for predicting Tc of clathrate hydrides. }
  \label{ml}
\end{figure*}

\subsection{\textbf{nn}}
\label{nn}
The fully connected neural network is an important algorithm in machine learning field that directly 
contributes to deep learning. It is theoretically possible to fit arbitrary functions with an appropriate number of 
layers and nodes and sufficiently large training sets, making it ideal for uncovering deeper relationships 
between factors and material properties and for abstracting design strategies for materials with target 
functionality. However, the realization of a neural network algorithm usually depends on the machine learning 
frameworks of Tensorflow or Pytorch that demands a high programming threshold. The FF7 code package 
equips an artificial-intelligence module “nn” that natively supports the command-line based interface for 
building and training artificial neural networks, which greatly reduces the operating difficulty. There is no 
need to worry about installing additional complex modules, just making sure that the NumPy library, which is 
the only third-party library that the “nn” module depends on, works properly. Users can easily build and 
train a three hidden layer neural network with 8, 16 and 8 nodes by command \\
\texttt{\$ ff7 nn train -hidden 8 16 8 -inX X.txt -label label.txt -trainrate 0.8 -batch 50 -epoch 20 -lr 0.01, }\\
where the parameter “-inX” and “-label” receive the files containing descriptors and target materials properties. 
The “-trainrate” parameter declares the ratio of the training set selected by the random strategy and the 
remaining parameters control the training of the neural network, e.g. the number of batches, epochs and the 
learning rate, respectively. The prediction error of each epoch and the finally training result will be summarized 
in figure “out.svg”. After training, the neural network model will be saved in file “model.npy” and the command\\
\texttt{\$ ff7 nn predict -model module.npy -predict predictX.txt}\\
is able to make properties predictions with trained models. All the input file of “nn” module share the same 
format with the first two lines and columns being the comment region that will be ignored when reading, which 
is fully compatible with data files saved by the “save” function of the “db” module.

\section{Example}
\label{example}
The realization of room temperature superconductivity is the long-sought goal of researchers. In this 
section, we demonstrated the power, user-friendliness and flexibility of FF7 by assisting the high-throughput 
calculations and superconducting properties analysis of compounds MXH$_{12}$ (M, X= Mg, Ca, Sc, Ti, Sr, Y, Zr, 
Ba, La, Ce, Hf, Th) at high pressure\cite{AE_MTP}. Firstly, we ran the command \\
\texttt{\$ ff7 gen -1 -2 [Mg,Ca,Sc,Ti,Sr,Y,Zr,Ba,La,Ce,Th]} \\
in an empty directory to construct a structure pool (i.e. the main working path in which the following 
commands were executed) containing compounds with stoichiometry MXH$_{12}$ generated by element 
substitution. Then, the high-throughput calculations for structural optimization and electronic band structures 
of all compounds were performed by the combination command 
\texttt{\$ ff7 htp bandos -e 600 -k 0.03 -dopt -p 200 -dbpath ./db\_dir/my.db}, \\
and the stoichiometries, structure files and electronic band structures files were summarized and saved in a 
database in the specified path (i.e. “./db\_dir”). This is the complete process of building a database through 
high-throughput calculations using the FF7 code package: only two lines of commands are required, which is 
quite intuitive and user-friendly. We can further build up a high-pressure structures pool by collecting stable 
compounds from previous high-pressure work and structure searching methods and construct high-pressure 
database with similar command. Users can browse the entire database by the command “\$ff7 db show -dbpath ./db\_dir”
or print the brief database information by adding a “-summary” flag. More commands for 
operating the database can be found in section 3.5. For the HTP calculation results of compounds MXH$_{12}$, we used the command \\
\texttt{\$ ff7 post ch -dbpath /home/HighPressure.db -byjobs} \\
and \\
\texttt{\$ ff7 post heatmap -M 1 -X 2 }\\
to calculate the formation enthalpy convex hulls based on the self-built high-pressure database and summarized 
them into a thermodynamic stability heatmap, as shown in Fig. \ref{ml}a. We then performed HTP calculations for 
their dynamical stability and electron-phonon coupling constant $\lambda$ using QE software by the command
\texttt{\$ ff7 phonon\_qe -e 80 -q 3 3 3 -k 12 12 12 -pps nc -dopt True} \\
Finally, we can acquire the phonon spectra diagrams and superconducting critical temperature (Tc) of stable compounds. 
The functions of HTP calculations and constructing database exhibited above were realized under the 
government of corresponding built-in “calculation cards”. Although FF7 code package equips thorough 
“calculation cards” and the calculation tasks for these preset scenarios can progress with excellent robustness, 
in real research scenarios of different users, the preset functions are not enough for various requirements. In 
this case, for deep understanding of the superconductivity of clathrate hydrides like MXH$_{12}$ and summarizing 
design rules for high-Tc hydrides, we may, for example, consider the related material features such as the 
contribution of H electrons to the total density of states at the fermi level ($HDos$) and the average H-H bound 
length ($l$). The FF7 code package provides an interface for customizing the “calculation card”. This code design 
makes it a solid secondary development platform allowing us to extremely extend FF7’s HTP calculations and 
database construction functionalities with minimal programming required. The example of customized 
“calculation card” and the post-processing script to extract the variable HDos and store it into the database are 
illustrated in Fig. \ref{ml}b (which for variable l are supplied in SM). In the “calculation card” shown in upper panel 
in Fig. \ref{ml}b, we assigned the variable “RunCommand” as “sleep 1” so that no DFT calculations are performed 
and declared “Dirname” as “dos” to ensure that the post-processing script is ran in the 
“.../structure\_pool/compound/dos” directory. After running the command\\
\texttt{\$ ff7 htp self -file Hdos.cc},\\
the database added a “Hdos” column and stored $HDos$ values of all the compounds in the structure pool. All in 
all, the programming design of “calculation\ card” could meet any specific requirement. 

The built-in machine learning module “nn” allows us to have a deep understanding of relationships 
between Tc and other material features and train a model for predicting Tc that facilitates further materials 
design. We used the command \\
\texttt{\$ ff7 db save -cols Hdos l -filename inX.txt}\\
to extract the descriptors of variable $RHdos$, $l$, $Hdos$ per H atom into a file “inX.txt” and command\\
\texttt{\$ ff7 db save -cols Tc -filename label.txt}\\
to obtain a “label.txt” file containing target property of Tc. Finally, we trained a fully connected neural network 
model built by two hidden layers with 4 nodes for predicting Tc of clathrate hydrides by the command\\
\texttt{\$ ff7 nn train -hidden 4 4 -inX X.txt -label label.txt -trainrate 0.8 -batch 6 -epoch 2500 -lr 0.0001}.\\
The loss function and the performance of the model are shown in Fig 6c and 6d, respectively. 
There is a strong correlation between them and this machine learning model can be used to predict the Tc of other clathrate hydrides.

\section{Conclusion}
\label{conclusion}
We herein introduce a self-develop code package named FF7 to assist in high-throughput DFT 
calculations and building user’s own database through a Linux command-line based interface. Mainstream 
DFT calculation software VASP and QE are both supported while the interactive interface remains fairly 
uniform for ease of use, and the high-throughput functions for the latter are groundbreaking. The design 
strategy of “calculation card” ensures flexibility of high-throughput calculations making the FF7 code package 
to be an easy-to-use programmable infrastructure that facilitates secondary development and build a robust 
and flexible connection between HTP calculations and the database enabling users to easily build databases 
that match their research interests. Full-featured post-processing modules with strong database connectivity 
are integrated into the FF7 code package to process data from high-throughput calculations, realize data 
visualization and generate formation enthalpy convex hulls, with the last being the highlight. As the heart of 
the FF7 code, the database module opens up a fully interactive interface to the user allowing complete mastery 
of the database. In particular, we developed a command-line based machine learning module that makes the 
process of building and training artificial neural networks as easy as “building blocks”, based on which we 
reveal the relationships between the $T_c$ of clathrate hydrides and other properties with low computational cost. 




\bibliographystyle{elsarticle-num}
\bibliography{ff7.bib}

\begin{thebibliography}{10}
\expandafter\ifx\csname url\endcsname\relax
  \def\url#1{\texttt{#1}}\fi
\expandafter\ifx\csname urlprefix\endcsname\relax\def\urlprefix{URL }\fi
\expandafter\ifx\csname href\endcsname\relax
  \def\href#1#2{#2} \def\path#1{#1}\fi

\bibitem{fourthpar}
A.~Agrawal, A.~Choudhary, Perspective: Materials informatics and big data: Realization of the “fourth paradigm” of science in materials science, APL Materials 4~(5) (2016) 053208.
\newblock \href {https://doi.org/10.1063/1.4946894} {\path{doi:10.1063/1.4946894}}.

\bibitem{prl_bigdata}
L.~M. Ghiringhelli, J.~Vybiral, S.~V. Levchenko, C.~Draxl, M.~Scheffler, Big data of materials science: Critical role of the descriptor, Phys. Rev. Lett. 114 (2015) 105503.
\newblock \href {https://doi.org/10.1103/PhysRevLett.114.105503} {\path{doi:10.1103/PhysRevLett.114.105503}}.

\bibitem{VASP}
G.~Kresse, J.~Furthm\"uller, Efficient iterative schemes for ab initio total-energy calculations using a plane-wave basis set, Phys. Rev. B 54 (1996) 11169--11186.
\newblock \href {https://doi.org/10.1103/PhysRevB.54.11169} {\path{doi:10.1103/PhysRevB.54.11169}}.

\bibitem{CASTEP}
M.~D. Segall, P.~J.~D. Lindan, M.~J. Probert, C.~J. Pickard, P.~J. Hasnip, S.~J. Clark, M.~C. Payne, First-principles simulation: ideas, illustrations and the castep code, Journal of Physics: Condensed Matter 14~(11) (2002) 2717.
\newblock \href {https://doi.org/10.1088/0953-8984/14/11/301} {\path{doi:10.1088/0953-8984/14/11/301}}.

\bibitem{QE}
P.~Giannozzi, S.~Baroni, N.~Bonini, M.~Calandra, R.~Car, C.~Cavazzoni, D.~Ceresoli, G.~L. Chiarotti, M.~Cococcioni, I.~Dabo, A.~Dal~Corso, S.~de~Gironcoli, S.~Fabris, G.~Fratesi, R.~Gebauer, U.~Gerstmann, C.~Gougoussis, A.~Kokalj, M.~Lazzeri, L.~Martin-Samos, N.~Marzari, F.~Mauri, R.~Mazzarello, S.~Paolini, A.~Pasquarello, L.~Paulatto, C.~Sbraccia, S.~Scandolo, G.~Sclauzero, A.~P. Seitsonen, A.~Smogunov, P.~Umari, R.~M. Wentzcovitch, Quantum espresso: a modular and open-source software project for quantum simulations of materials, Journal of Physics: Condensed Matter 21~(39) (2009) 395502.
\newblock \href {https://doi.org/10.1088/0953-8984/21/39/395502} {\path{doi:10.1088/0953-8984/21/39/395502}}.

\bibitem{CGCNN}
T.~Xie, J.~C. Grossman, Crystal graph convolutional neural networks for an accurate and interpretable prediction of material properties, Phys. Rev. Lett. 120 (2018) 145301.
\newblock \href {https://doi.org/10.1103/PhysRevLett.120.145301} {\path{doi:10.1103/PhysRevLett.120.145301}}.

\bibitem{ALIGNN}
K.~Choudhary, B.~DeCost, Atomistic line graph neural network for improved materials property predictions, npj Computational Materials 8 (2022) 221.
\newblock \href {https://doi.org/10.1038/s41524-022-00913-5} {\path{doi:10.1038/s41524-022-00913-5}}.

\bibitem{MGI}
Materials genome initiative for global competitiveness, OSTP (June 2011).

\bibitem{pymatgen_htp}
A.~Jain, G.~Hautier, C.~J. Moore, S.~{Ping Ong}, C.~C. Fischer, T.~Mueller, K.~A. Persson, G.~Ceder, A high-throughput infrastructure for density functional theory calculations, Computational Materials Science 50~(8) (2011) 2295--2310.
\newblock \href {https://doi.org/https://doi.org/10.1016/j.commatsci.2011.02.023} {\path{doi:https://doi.org/10.1016/j.commatsci.2011.02.023}}.

\bibitem{MP}
A.~Jain, S.~P. Ong, G.~Hautier, W.~Chen, W.~D. Richards, S.~Dacek, S.~Cholia, D.~Gunter, D.~Skinner, G.~Ceder, K.~A. Persson, Commentary: The materials project: A materials genome approach to accelerating materials innovation, APL Materials 1~(1) (2013) 011002.
\newblock \href {https://doi.org/10.1063/1.4812323} {\path{doi:10.1063/1.4812323}}.

\bibitem{ICSD}
D.~Zagorac, H.~M{\"{u}}ller, S.~Ruehl, J.~Zagorac, S.~Rehme, {Recent developments in the Inorganic Crystal Structure Database: theoretical crystal structure data and related features}, Journal of Applied Crystallography 52~(5) (2019) 918--925.
\newblock \href {https://doi.org/10.1107/S160057671900997X} {\path{doi:10.1107/S160057671900997X}}.

\bibitem{OQMD}
S.~Kirklin, J.~E. Saal, B.~Meredig, A.~Thompson, J.~W. Doak, M.~Aykol, S.~Rühl, C.~Wolverton, The open quantum materials database (oqmd): assessing the accuracy of dft formation energies, npj Computational Materials 1 (2015) 15010.
\newblock \href {https://doi.org/10.1038/npjcompumats.2015.10} {\path{doi:10.1038/npjcompumats.2015.10}}.

\bibitem{JAMIP}
X.-G. Zhao, K.~Zhou, B.~Xing, R.~Zhao, S.~Luo, T.~Li, Y.~Sun, G.~Na, J.~Xie, X.~Yang, X.~Wang, X.~Wang, X.~He, J.~Lv, Y.~Fu, L.~Zhang, Jamip: an artificial-intelligence aided data-driven infrastructure for computational materials informatics, Science Bulletin 66~(19) (2021) 1973--1985.
\newblock \href {https://doi.org/https://doi.org/10.1016/j.scib.2021.06.011} {\path{doi:https://doi.org/10.1016/j.scib.2021.06.011}}.

\bibitem{hardness}
Y.-J. Chang, C.-Y. Jui, W.-J. Lee, A.-C. Yeh, Prediction of the composition and hardness of high-entropy alloys by machine learning, JOM 71 (2019) 3433--3442.
\newblock \href {https://doi.org/10.1007/s11837-019-03704-4} {\path{doi:10.1007/s11837-019-03704-4}}.

\bibitem{PRM_mlTc}
H.~Tran, T.~N. Vu, Machine-learning approach for discovery of conventional superconductors, Phys. Rev. Mater. 7 (2023) 054805.
\newblock \href {https://doi.org/10.1103/PhysRevMaterials.7.054805} {\path{doi:10.1103/PhysRevMaterials.7.054805}}.

\bibitem{PRB2023_mlTc}
A.~D. Smith, S.~B. Harris, R.~P. Camata, D.~Yan, C.-C. Chen, Machine learning the relationship between debye temperature and superconducting transition temperature, Phys. Rev. B 108 (2023) 174514.
\newblock \href {https://doi.org/10.1103/PhysRevB.108.174514} {\path{doi:10.1103/PhysRevB.108.174514}}.

\bibitem{Hutcheon}
M.~J. Hutcheon, A.~M. Shipley, R.~J. Needs, Predicting novel superconducting hydrides using machine learning approaches, Phys. Rev. B 101 (2020) 144505.
\newblock \href {https://doi.org/10.1103/PhysRevB.101.144505} {\path{doi:10.1103/PhysRevB.101.144505}}.

\bibitem{Shipley}
A.~M. Shipley, M.~J. Hutcheon, R.~J. Needs, C.~J. Pickard, High-throughput discovery of high-temperature conventional superconductors, Phys. Rev. B 104 (2021) 054501.
\newblock \href {https://doi.org/10.1103/PhysRevB.104.054501} {\path{doi:10.1103/PhysRevB.104.054501}}.

\bibitem{sa_luo}
Z.~Luo, W.~Gao, Q.~Jiang, Determinants of vacancy formation and migration in high-entropy alloys, Science Advances 11~(1) (2025) eadr4697.
\newblock \href {http://arxiv.org/abs/https://www.science.org/doi/pdf/10.1126/sciadv.adr4697} {\path{arXiv:https://www.science.org/doi/pdf/10.1126/sciadv.adr4697}}, \href {https://doi.org/10.1126/sciadv.adr4697} {\path{doi:10.1126/sciadv.adr4697}}.

\bibitem{VASPKIT}
V.~Wang, N.~Xu, J.-C. Liu, G.~Tang, W.-T. Geng, Vaspkit: A user-friendly interface facilitating high-throughput computing and analysis using vasp code, Computer Physics Communications 267 (2021) 108033.
\newblock \href {https://doi.org/https://doi.org/10.1016/j.cpc.2021.108033} {\path{doi:https://doi.org/10.1016/j.cpc.2021.108033}}.

\bibitem{QVASP}
W.~Yi, G.~Tang, X.~Chen, B.~Yang, X.~Liu, qvasp: A flexible toolkit for vasp users in materials simulations, Computer Physics Communications 257 (2020) 107535.
\newblock \href {https://doi.org/https://doi.org/10.1016/j.cpc.2020.107535} {\path{doi:https://doi.org/10.1016/j.cpc.2020.107535}}.

\bibitem{VASPMATE}
Z.~Pan, Z.~Liu, T.~Xu, D.~Legut, R.~Zhang, Vaspmate: An integrated user-interface program for high-throughput first principles computations through vasp code, Computational Materials Science 233 (2024) 112707.
\newblock \href {https://doi.org/https://doi.org/10.1016/j.commatsci.2023.112707} {\path{doi:https://doi.org/10.1016/j.commatsci.2023.112707}}.

\bibitem{NP_H3S}
M.~Einaga, M.~Sakata, T.~Ishikawa, K.~Shimizu, M.~I. Eremets, A.~P. Drozdov, I.~A. Troyan, N.~Hirao, Y.~Ohishi, Crystal structure of the superconducting phase of sulfur hydride, Nature Physics 12 (2016) 835--838.
\newblock \href {https://doi.org/10.1038/nphys3760} {\path{doi:10.1038/nphys3760}}.

\bibitem{N_H3S}
A.~P. Drozdov, M.~I. Eremets, I.~A. Troyan, V.~Ksenofontov, S.~I. Shylin, Conventional superconductivity at 203 kelvin at high pressures in the sulfur hydride system, Nature 525~(7567) (2015) 73--76.
\newblock \href {https://doi.org/10.1038/nature14964} {\path{doi:10.1038/nature14964}}.

\bibitem{ddfH3S}
D.~Duan, Y.~Liu, F.~Tian, D.~Li, X.~Huang, Z.~Zhao, H.~Yu, B.~Liu, W.~Tian, T.~Cui, Pressure-induced metallization of dense (h2s)2h2 with high-tc superconductivity, Scientific Reports 4~(1) (2014) 6968.
\newblock \href {https://doi.org/10.1038/srep06968} {\path{doi:10.1038/srep06968}}.

\bibitem{pnas_CaH6}
H.~Wang, J.~S. Tse, K.~Tanaka, T.~Iitaka, Y.~Ma, Superconductive sodalite-like clathrate calcium hydride at high pressures, Proceedings of the National Academy of Sciences 109~(17) (2012) 6463--6466.
\newblock \href {https://doi.org/10.1073/pnas.1118168109} {\path{doi:10.1073/pnas.1118168109}}.

\bibitem{an_Ca4H23}
D.~An, D.~Duan, Z.~Zhang, Q.~Jiang, T.~Ma, Z.~Huo, H.~Song, T.~Cui, Type-i clathrate calcium hydride and its hydrogen-vacancy structures at high pressure, Phys. Rev. B 110 (2024) 054505.
\newblock \href {https://doi.org/10.1103/PhysRevB.110.054505} {\path{doi:10.1103/PhysRevB.110.054505}}.

\bibitem{ma_CaH6}
L.~Ma, K.~Wang, Y.~Xie, X.~Yang, Y.~Wang, M.~Zhou, H.~Liu, X.~Yu, Y.~Zhao, H.~Wang, G.~Liu, Y.~Ma, High-temperature superconducting phase in clathrate calcium hydride ${\mathrm{cah}}_{6}$ up to 215 k at a pressure of 172 gpa, Phys. Rev. Lett. 128 (2022) 167001.
\newblock \href {https://doi.org/10.1103/PhysRevLett.128.167001} {\path{doi:10.1103/PhysRevLett.128.167001}}.

\bibitem{exp_LaBeH8}
Y.~Song, J.~Bi, Y.~Nakamoto, K.~Shimizu, H.~Liu, B.~Zou, G.~Liu, H.~Wang, Y.~Ma, Stoichiometric ternary superhydride ${\mathrm{labeh}}_{8}$ as a new template for high-temperature superconductivity at 110 k under 80 gpa, Phys. Rev. Lett. 130 (2023) 266001.
\newblock \href {https://doi.org/10.1103/PhysRevLett.130.266001} {\path{doi:10.1103/PhysRevLett.130.266001}}.

\bibitem{pre_LaBeH8}
Z.~Zhang, T.~Cui, M.~J. Hutcheon, A.~M. Shipley, H.~Song, M.~Du, V.~Z. Kresin, D.~Duan, C.~J. Pickard, Y.~Yao, Design principles for high-temperature superconductors with a hydrogen-based alloy backbone at moderate pressure, Phys. Rev. Lett. 128 (2022) 047001.
\newblock \href {https://doi.org/10.1103/PhysRevLett.128.047001} {\path{doi:10.1103/PhysRevLett.128.047001}}.

\bibitem{AE_MTP}
T.~Ma, Z.~Zhang, M.~Du, Z.~Huo, W.~Chen, F.~Tian, D.~Duan, T.~Cui, High-throughput calculation for superconductivity of sodalite-like clathrate ternary hydrides {MXH}$_{12}$ at high pressure, Materials Today Physics 38 (2023) 101233.
\newblock \href {https://doi.org/https://doi.org/10.1016/j.mtphys.2023.101233} {\path{doi:https://doi.org/10.1016/j.mtphys.2023.101233}}.

\end{thebibliography}







\end{document}